\documentclass{article}

\usepackage{natbib}
\setcitestyle{numbers}
\usepackage[english]{babel}
\usepackage{array}
\usepackage{float}
\usepackage{url}

\usepackage[letterpaper,top=2cm,bottom=2cm,left=3cm,right=3cm,marginparwidth=1.75cm]{geometry}

\usepackage{amsmath}
\usepackage{graphicx}
\usepackage[colorlinks=true, allcolors=blue]{hyperref}
\usepackage[dvipsnames]{xcolor}

\title{Using Physics Simulations to Find Targeting Strategies in Competitive Bowling}
\author{Simon Ji, Shouzhuo Yang, Wilber Dominguez, Cacey Bester}

\begin{document}
\maketitle

\begin{abstract}
This article demonstrates a new approach to finding ideal bowling targeting strategies through computer simulation. To model bowling ball behaviour, a system of 5 coupled differential equations is derived using Euler’s equations for rigid body rotations. We used a computer program to demonstrate the phases of ball motion and output a plot that displays the optimum initial conditions that can lead to a strike. When the bowler is modeled to be imperfect, it is shown that some targeting strategies can lead to higher strike rates due to the “miss room” created by the inhomogeneity of the oil pattern.
\end{abstract}

\section{Introduction}

Bowling remains as one of the most popular sports in the U.S. , with over 45 million people participating regularly in bowling as of 2017. \cite{Outdoor} With millions of dollars at stake every year in national competitions, significant research has been done to understand how players can achieve higher scores. Due to the complexity of the calculations and the vast number of variables that can affect the ball’s trajectory, most of the research has focused on statistical analysis from empirical data instead of theoretical modeling. For example, the 2018 U.S. Bowling Congress (USBC) Equipment Specifications Report used “37 bowlers with a full range of revolutions per minute (RPM) rates” rather than a computer model.\cite{EquipmentStudy} 

Literature on quantitative analysis of bowling physics is rare due to the many parameters involved, but has been attempted by Frohlich, Hopkins and Huston over the past few decades.\cite{Frohlich}\cite{Hopkins}\cite{Huston} Frohlich and Huston created mathematical models that took into account the effects of a weighted core within the bowling ball, and provided simulation results for a small sample of parameter values, including effects of varying radius of gyration (RG), center of gravity (CG) offset from the geometric center of ball and initial angular velocity. The simulations demonstrated the qualitative effects of changing certain variables, but only assumed simple friction profiles.

The purpose of this paper is to demonstrate bowling target strategies through a simulation that samples a large number of possible initial conditions, and explores the effects of realistic friction profiles based on modern competition oil patterns. The program can then provide the user with the best possible starting positions. These results would be much easier to obtain compared to empirical methods, and will be useful for both competitors and tournament oil pattern designers, who would quickly find the ideal way to tackle the lane condition. We verify the success of the simulation by testing the likelihood of a strike with changing entry positions and lane oil patterns. 

\pagebreak

\section{Methods}

\subsection{Equations of Motion}

The equations of motion derived in this section describe rigid body rotation using a rotating frame of reference fixed to the ball. The derivations use a novel approach through Euler's equations, as they are only dependent on the principal moments of inertia, and not the non-diagonal terms in the inertia tensor; this is important as the full inertia tensor of a reference frame fixed to space cannot be easily determined simply based on the published Radius of Gyration (RG) and differential values of each ball. Previous theories derived by Hopkins assumed a perfectly uniform, spherical ball \cite{Hopkins}, while theories from Frohlich required knowing the non-diagonal terms in the inertia tensor of the bowling ball. \cite{Frohlich} The CG offset (distance from the center of gravity to the geometric center of the ball, must be below 1mm to be competition-legal \cite{Frohlich}) is assumed to be zero for this study in order to simplify calculations.

\begin{table}[H]
\centering
\begin{tabular}{ |m{8em}|m{30em}| }
\hline Parameters & Explanation \\\hline
\(v_0, \theta_0\) & Initial velocity and angle of the ball. \(\theta_0\) is measured relative to the y-axis, typically ranging from 0 to 5 degrees.\\\hline
\(\omega_{x,0}, \omega_{y,0}, \omega_{z,0}\) & Initial angular velocities imparted on the ball. A skilled bowler typically generates around 300-400 revolutions per minute \cite{Frohlich}, while some bowlers using the two-handed style are able to generate up to 600 revolutions per minute. The direction of rotation varies depending on bowling style, and is typically determined using measurements of “positive axis point”.\\\hline
\(I_{x'}, I_{y'}, I_{z'}\) & The principal moments of inertia, measured in a rotated frame so that the highest inertia occurs at the y’ axis. These values are different depending on the bowling ball, and can be derived using three values given by the manufacturer: Radius of Gyration (RG), Differential (Diff), Intermediate Differential (Int.\,Diff). Different combinations lead to different ball trajectories.\newline
$I_{y'} = m ($RG$)^2$\newline
$I_{x'} = m ($RG+Diff$)^2$\newline
$I_{z'} = m ($RG+Diff+Int.\,Diff$)^2$
\\\hline
\(\mu\) & Kinetic friction coefficient between the ball’s surface and the lane. When the lane is completely dry, this value is roughly 0.2. When oil is applied to the lanes, \(\mu\) decreases to around 0.04. \cite{Banerjee} Exact values will depend on the lane conditions, oil thickness and ball surface characteristics. \\\hline
\(\phi\) & Angle between the major moments of inertia axis y’ and the lane axis y. (See figure 1) This value would change based on the bowler’s style and the way finger holes are drilled into the ball.\\\hline
\(x_0, y_0\) & Starting Position of the ball. \(x_0\) is easily adjusted by the bowler; \(y_0\) is typically close to zero; however, for rare cases when the lane friction is too high, professional bowlers sometimes choose to “loft” the ball and increase \(y_0\) by 3 to 5 meters. This allows a lower total contact time with the lane, hence decreasing hook.\\\hline
m & Mass of bowling ball. Most league bowlers use masses between 6.3 to 7.3 kg (14 to 16 lbs). \\\hline
r & Radius of bowling ball. A typical ball has a radius of 10.85 cm. \\\hline

\end{tabular}
\caption{\label{tab:widgets}List of parameters used to derive the ball trajectory.}
\end{table}

\begin{figure}[H]
\centering
\includegraphics[width=1\textwidth]{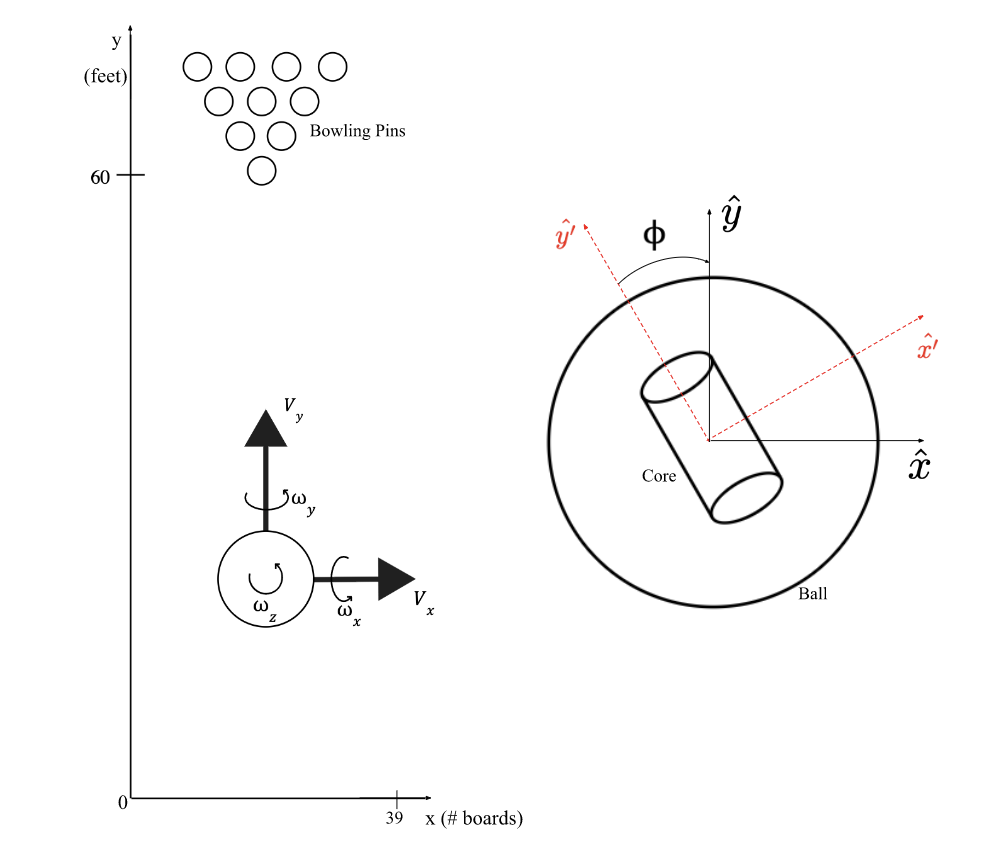}
\caption{\label{fig:1}Axis Definitions with respect to the lane and the bowling ball. The x-axis of a bowling lane is typically measured in boards. A USBC-approved bowling lane has 39 boards, each measuring 2.74cm. The y’ axis is aligned with the minimum moment of inertia axis of the core.}
\end{figure}

Ignoring air resistance, the only force of interaction between the ball and the lane is at the contact surface. The direction of the friction force on the ball is therefore purely dependent on the contact surface velocity \(\Vec{v^b}\), as illustrated in figure 2.

\begin{figure}[H]
\centering
\includegraphics[width=1\textwidth]{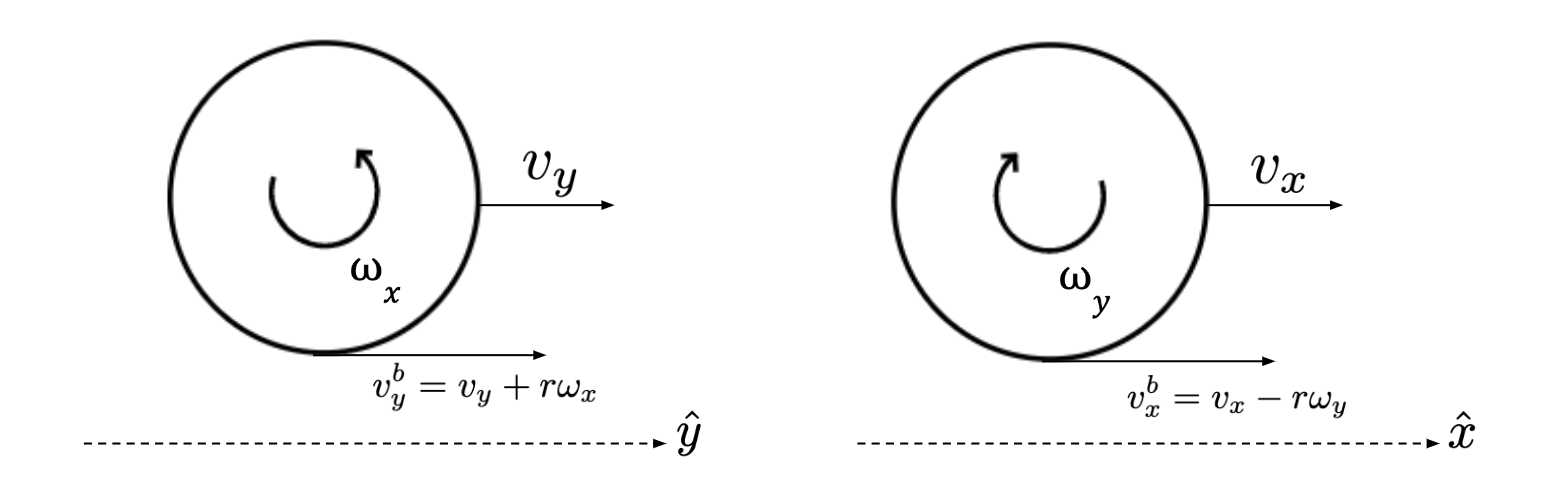}
\caption{\label{fig:2}Surface interactions between the ball and the lane along the \(\hat{x}\) and \(\hat{y}\) directions.}
\end{figure}
\pagebreak
\begin{equation}
v_x^b = v_x - r\omega_y 
\end{equation}
\begin{equation}
v_y^b = v_y + r\omega_x
\end{equation}

The surface speed at the ball’s point of contact with the lane can therefore be calculated using the formula
\begin{equation}
v^b = \sqrt{(v_x^b)^2+(v_y^b)^2} = \sqrt{(v_x-r\omega_y)^2+(v_y+r\omega_x)^2}.
\end{equation}

Friction forces in the \(\hat{x}\) and \(\hat{y}\) directions respectively equal to
\begin{equation}
F_x = m\dot{v}_x = -\mu mg\frac{v_x^b}{v^b}
\end{equation}
\begin{equation}
F_y = m\dot{v}_y = -\mu mg\frac{v_y^b}{v^b}.
\end{equation}

The friction force then applies a torque to the ball. Zero torque is applied in the z-direction, as we assume the ball contacts the surface at a single point.
\begin{equation}
\Vec{\tau} = \Vec{r} \times \Vec{F} = \begin{bmatrix}0\\0\\-r\end{bmatrix} \times \begin{bmatrix}m\dot{v}_x\\m\dot{v}_y\\0\end{bmatrix} = \begin{bmatrix}rm\dot{v}_y\\rm\dot{v}_x\\0\end{bmatrix}
\end{equation}

Combining the equations above with Euler’s Equations for a Rigid body, in the ball core’s frame,

\begin{equation}
\tau_{x'} = I_{x'}\dot{\omega}_{x'} - (I_{y'}-I_{z'})\omega_{y'}\omega_{z'}
\end{equation}
\begin{equation}
\tau_{y'} = I_{y'}\dot{\omega}_{y'} - (I_{z'}-I_{x'})\omega_{z'}\omega_{x'}
\end{equation}
\begin{equation}
\tau_{z'} = I_{z'}\dot{\omega}_{z'} - (I_{x'}-I_{y'})\omega_{x'}\omega_{y'}
\end{equation}

5 first-order, coupled differential equations are derived for the terms \(v_x, v_y, \omega_x, \omega_y, \omega_z\).
\begin{equation}
\dot{v_x} = -\mu g\frac{v_x-r\omega_y}{\sqrt{(v_x-r\omega_y)^2+(v_y+r\omega_x)^2}}
\end{equation}
\begin{equation}
\dot{v_y} = -\mu g\frac{v_y+r\omega_x}{\sqrt{(v_x-r\omega_y)^2+(v_y+r\omega_x)^2}}
\end{equation}
\begin{equation}
\dot{\omega_x}\cos{\phi}+\dot{\omega_y}\sin{\phi} = \frac{1}{I_{x'}}((\dot{v}_y\cos{\phi}+\dot{v}_x\sin{\phi})+(I_{y'}-I_{z'})(-\omega_x\sin{\phi}+\omega_y\cos{\phi})\omega_z)
\end{equation}
\begin{equation}
-\dot{\omega_x}\sin{\phi}+\dot{\omega_y}\cos{\phi} = \frac{1}{I_{y'}}((-\dot{v}_y\sin{\phi}-\dot{v}_x\cos{\phi})+(I_{z'}-I_{x'})(\omega_x\cos{\phi}+\omega_y\sin{\phi})\omega_z)
\end{equation}
\begin{equation}
\dot{\omega}_z = \frac{I_{x'}-I_{y'}}{I_z}(\omega_x\cos{\phi}+\omega_y\sin{\phi})(-\omega_x\sin{\phi}+\omega_y\cos{\phi})
\end{equation}

For equations 12 to 14, an axis rotation of degree \(\phi\) must be done on \(\omega_{x,y,z}\) and \(v_{x,y,z}\), as principal moments of inertia are measured in the ball's reference frame (see figure 1).

\subsection{Oil Patterns and Strike Chance}

In competitive bowling, oil is laid out on the lanes in patterns specifically designed to create challenging friction profiles that leave little room for error. Sample oil patterns are illustrated in figure 3. During competition, bowlers have to make educated guesses on ball choice, targeting and ball speed based to maximize their strike chance. 

\begin{figure}[H]
\centering
\includegraphics[width=0.5\textwidth]{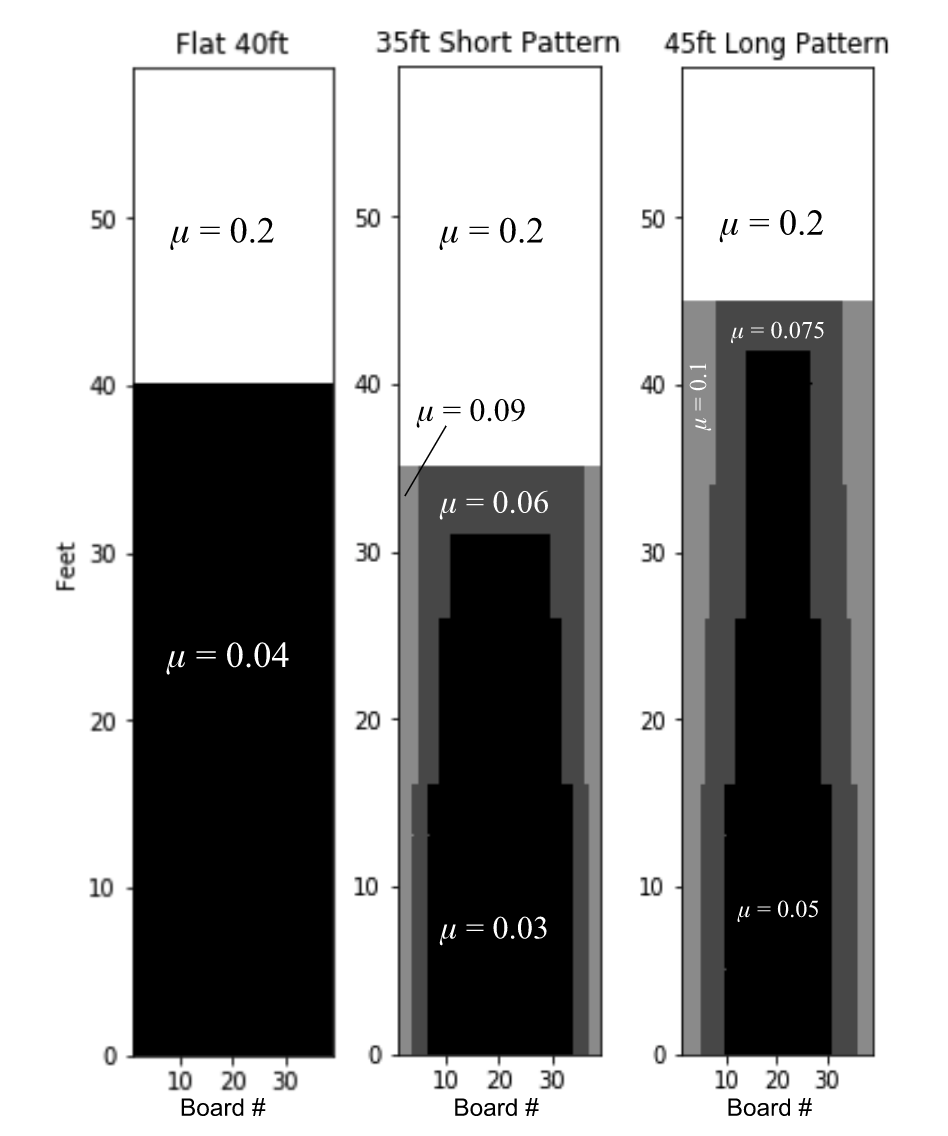}
\caption{\label{fig:3}Samples of typical competition oil patterns. Pattern lengths below 38ft (11.6m) are considered “short” and above 43ft (13.1m) are considered “long”. Friction coefficients for the flat pattern are estimates based on values given in Banerjee. \cite{Banerjee} The 35ft (10.7m) and 45ft (13.7m) patterns are rough approximations of PBA cheetah and shark patterns respectively, with their \(\mu\) values based on the total volume of oil laid onto the lanes.}
\end{figure}

Two major factors contribute to the chance of a strike: entry position and entry angle. Entry position is calculated based on the ball’s position at the moment it makes impact with the pins (y=60 feet). Entry angle is the angle that the ball trajectory makes relative to the vertical at y=60 feet. The ideal position of entry is between 1.3 to 14.0 cm (0.5-5.5 in) offset from the lane center, typically referred by bowlers as the “pocket”. A 2009 empirical study conducted by USBC provided the strike percentages at various entry angles and positions.\cite{PinCarry} This data is incorporated into the simulation, allowing it to calculate a score based on the ball’s trajectory.

\begin{figure}[H]
\centering
\includegraphics[width=0.5\textwidth]{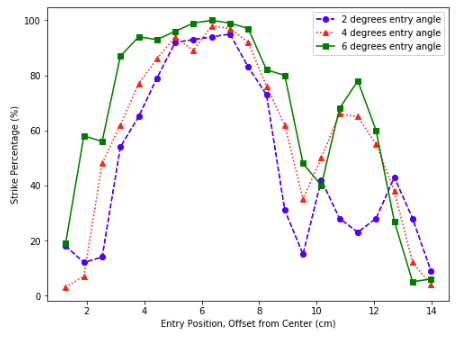}
\caption{\label{fig:4}Graph of strike percentages at 2,4,6 degree entry angles respectively, as a function on entry position. Data taken from the 2009 Bowl Expo presentation. \cite{PinCarry} Total strike percentage is highest at 6 degrees and lowest at 2 degrees.}
\end{figure}

\subsection{Miss Room for an Imperfect Bowler}

In real life, no bowlers can hit their targets with 100 percent accuracy. The best professional players can get to within around 0.1 degrees from their target initial angle, which can correspond to a difference of a few centimeters at the other end of the lane. As shown in figure 4, this change will significantly affect the chances of a strike. In this study, the inaccuracy is modeled using a Gaussian distribution of starting angles, with the standard deviation equal to 0.1 degrees.

\section{Simulation Results}

All simulation results are completed using the following initial conditions: \(y_0\)=0m, \(\omega_{x,0}=\)-30rad/s, \(\omega_{y,0}\)=-30rad/s, \(\omega_{z,0}\)=10rad/s, RG=6.35cm, Diff=0.1cm, Int. Diff=0cm, m=6.8kg, \(v_0\)=8m/s. These values are typical for a competitive bowler. \cite{Frohlich} \(x_0\) and \(\theta_0\) are varied to find the different strike results.

\subsection{Phases of Ball Motion}

The ball’s motion on the lane can be categorized into two phases: sliding and rolling. During the sliding phase, the contact point between the ball and lane is not stationary in the lane's frame of reference (\(v_b\neq0\)). Due to the low friction and large rotational inertia imparted by the bowler at the start, the ball spends most of its travel path in the sliding phase. Once the surface speed decreases to zero, pure rolling phase begins as the contact point is now stationary relative to the lane. The ball travels at a straight line towards the pins as no more torque is applied. Figure 5 demonstrates this through a plot of \(v^b\) over time on a flat oil pattern.

\begin{figure}[H]
\centering
\includegraphics[width=0.8\textwidth]{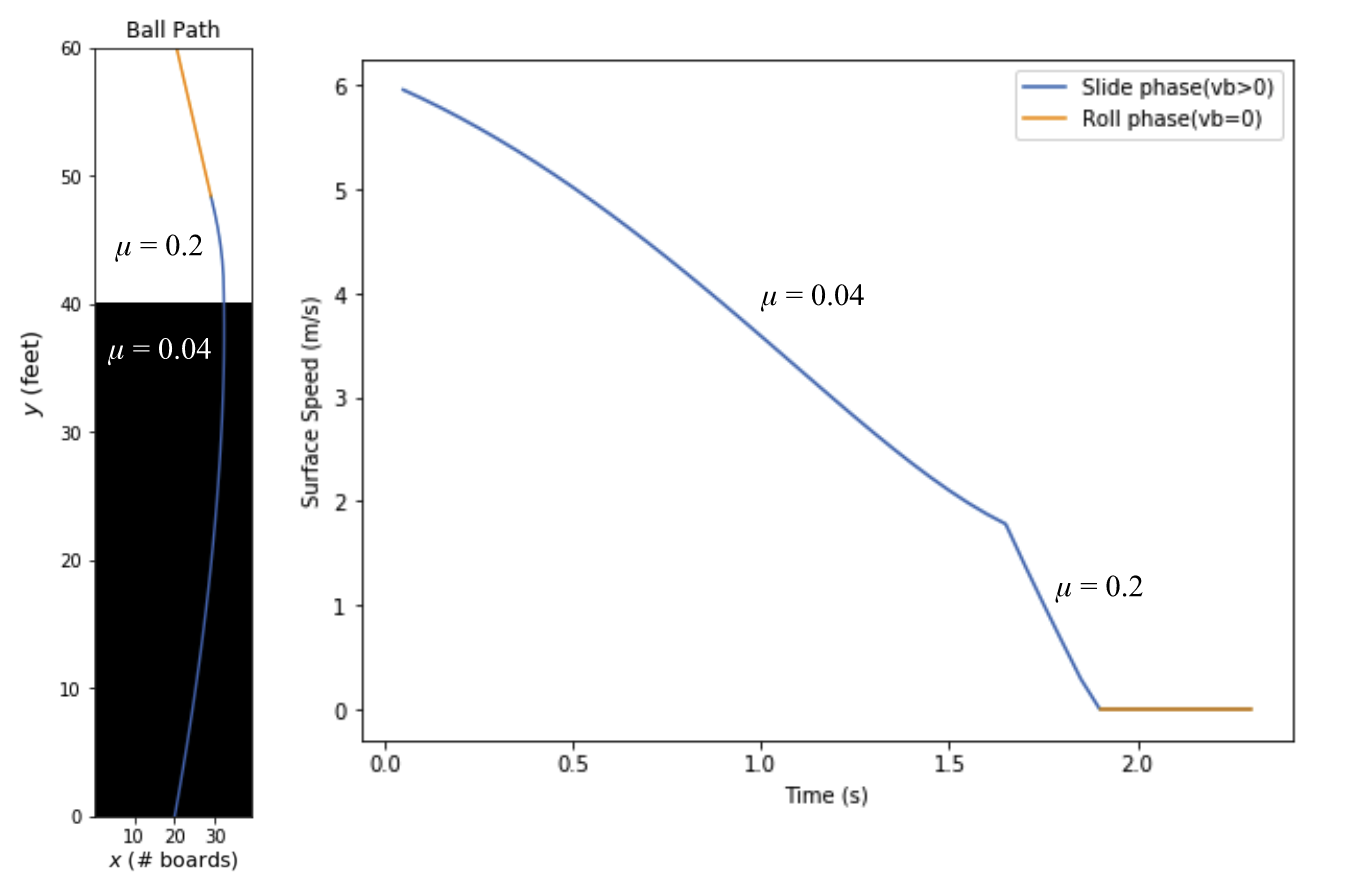}
\caption{\label{fig:5} Ball’s surface speed over time, and the corresponding locations in the ball path, starting at board 20 with initial angle of 3 degrees on the 40ft flat oil pattern. The rate of change of surface speed is closely related to the friction coefficient.}
\end{figure}
\pagebreak

\subsection{Finding the Ideal Targeting Strategy}
In order to find the starting position and angles that can lead to a strike, we calculated the outcomes for all possible \(x_0\) values (board 0-39) and starting angle \(\theta_0\) between 0-6 degrees relative to the y-axis. Figure 6 shows the results of this simulation for the flat pattern:

\begin{figure}[H]
\centering
\includegraphics[width=1\textwidth]{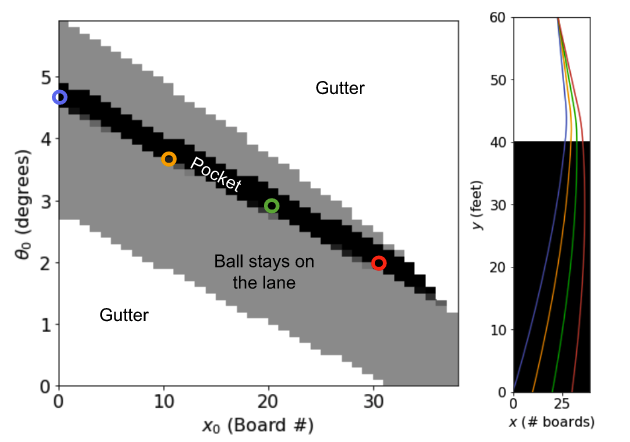}
\caption{\label{fig:6} 2D plot of simulation results, at all possible $x_0$ values and \(\theta_0\) between 0-6$^{\circ}$ relative to the y-axis. The black pixels correspond to a pocket hit (1-14 cm off-center at entry). The gray areas correspond to when the ball does not go into the gutter, but does not hit the pocket area. A sample of pocket hit paths are shown in the right figure.}
\end{figure}

The simulation results can then be combined with the strike chance plots in figure 4, to determine the exact chances for these “pocket hits” in the black region to result in a strike. In order to incorporate the “imperfect bowler” idea from the previous section, a weighted average is calculated based on a Gaussian distribution of starting angles. 

\pagebreak

\begin{figure}[H]
\centering
\includegraphics[width=1\textwidth]{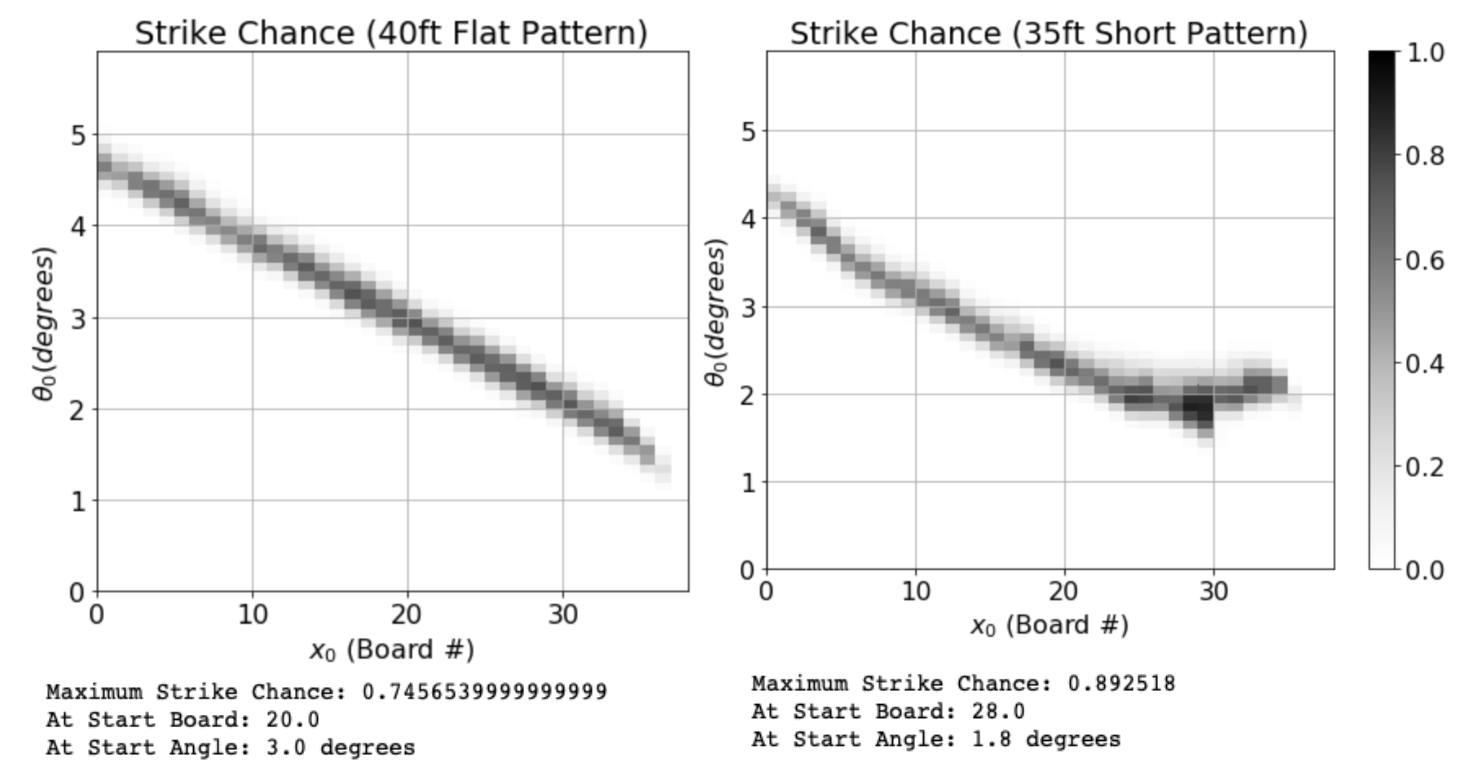}
\caption{\label{fig:7} Strike chance plots for flat and short patterns, with the best starting positions provided by the program. Flat pattern exhibited a linear profile, while the short pattern exhibited a lopsided “v-shape” due to inhomogeneity of the oil pattern.}
\end{figure}

Figure 7 shows that there does not seem to be a strong preference towards any starting position for the flat pattern, with only a narrow region of possible starting conditions. For the short pattern, however, a much wider area of viable starting positions can be found to have high strike percentage near the turning point of the v-shape, corresponding to the inhomogeneous oil distribution near the edge of the lane. The highest strike chance is also found in this region, with an estimated value of 0.89, much larger than 0.75 found in the flat pattern. We define the "ideal line" on the short pattern to be the starting board and angle found by the simulation: $x_0=28.0$ boards, $\theta_0=1.8^{\circ}$.

\begin{figure}[H]
\centering
\includegraphics[width=0.5\textwidth]{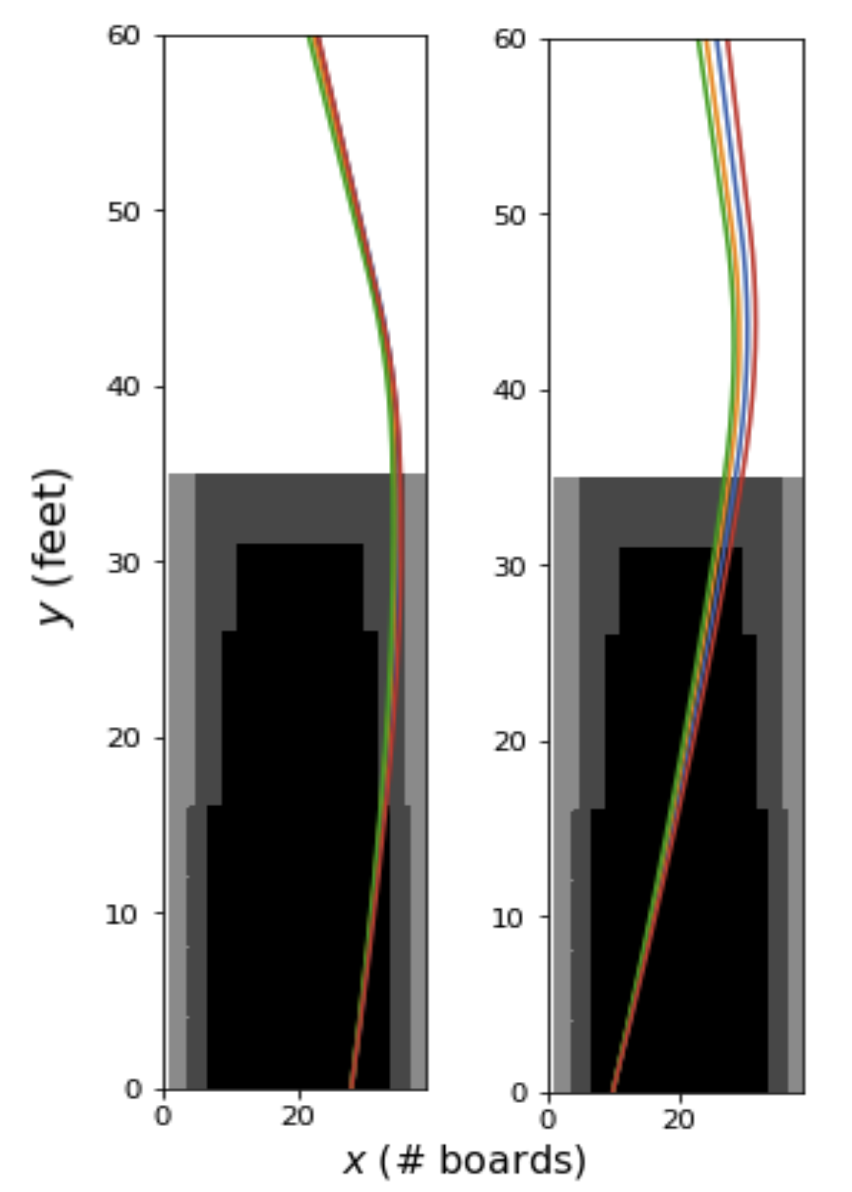}
\caption{\label{fig:8} Comparison between an “ideal line” ($x_0=28$ boards, $\theta_0=1.8^{\circ}$) and a “non-ideal line” ($x_0=10$ boards, $\theta_0=3.2^{\circ}$). The plotted paths represent small deviations ($\Delta\theta_0 = \pm 0.05^{\circ}, \pm 0.15^{\circ}$) to the original input.}
\end{figure}

The significance of the “maximum strike chance” is more clearly shown in figure 8. Even though both the “ideal line” ($x_0=28$ boards, $\theta_0=1.8^{\circ}$) and the “non-ideal line” ($x_0=10$ boards, $\theta_0=3.2^{\circ}$) lead to a pocket hit, small deviations in starting angle due to human error results in a much larger deviations in the final entry position for the “non-ideal line” (8 cm, compared to less than 3 cm for the ideal line).

The ideal line forms a path that closely aligns with the boundary between two friction zones of the short oil pattern. This makes sense qualitatively: if the bowler misses to the right, the higher friction near the gutter would accelerate the ball to the left; similarly, the lower friction in the center means a shot that misses left will not hook early. While this result is not surprising to experienced bowlers, new bowlers often struggle to understand this concept and make adjustments that leave them with a much smaller miss room.

\pagebreak
\section{Discussion}

\subsection{Improvements to the Model}
In the theory derivations, several important factors that impact the ball’s motion could be considered for future study. These factors include:

\begin{enumerate}
\item CG Offset: The actual center of mass for the bowling ball can be slightly offset from the geometrical center, due to manufacturing defects and materials lost during finger hole drilling. USBC regulations specify that it cannot be more than 1mm offset from center. The effect of a mass offset has been quantified in previous studies, which showed that even a 1mm offset will lead to a measurable change in the resulting entry board.
\item Lane topography: While bowling lanes are installed to be as flat as possible, small deviations in the lane topography exist that can affect the paths of a bowling ball. This is especially noticeable for harder flat patterns where its effects are not masked by the varied friction coefficients. USBC has developed methods to accurately measure the deviations, and some high-profile competitions even give out booklets showing the topography for each lane for the competitors to use as reference. 
\item Coverstock: The outer surface of a bowling ball directly affects the friction coefficient the ball experiences. Many different “coverstocks” have been developed to create useful hook profiles, such as reactive resin covers that have microscopic pores to absorb oil and increase the friction. Additionally, bowlers are allowed to apply sandpaper before a tournament to fine tune the surface property depending on the lane conditions.
\item Surface hardness: The current theories model the surface interactions by assuming that contact area is infinitesimally small, as the ball is infinitely hard. In reality, deformations of the ball and the lane would result in a loss of rotational energy, causing velocities to continually decrease even in the rolling phase. A torque would also be applied in the z-direction, as the ball no longer only contacts the surface at a single point on the z-axis.
\item Oil breakdown: As the ball travels down the lane, the oil pattern on the lane will be affected; some oil will be transferred from the lane to the ball’s surface. Depending on the ball’s cover type, the oil will either be absorbed or deposited at another location on the lane. These changes significantly affect the targeting and ball choice, and will need to be accounted for in order to make accurate predictions after a few games are played. 

\end{enumerate}

Realistically, it would be very difficult to incorporate all of these factors into the same program, as most of them require tedious and expensive measurements that are only accessible in the most advanced bowling alleys and training centers. The program would therefore only act as a rough estimation to help bowlers choose the balls they would bring to the competition, as well as allowing oil pattern designers to determine if they created a profile that is challenging but also fun to play in. 

However, in rare cases such as high-stakes televised finals, all of these parameters can be controlled. Measurements of the topography, friction coefficients and bowling ball properties can be made ahead of time. During the game, SPECTO systems are able to provide the exact trajectories from each shot, allowing an accurate prediction of oil breakdown. A well-calibrated program would then be able to provide bowlers with the best possible strategies and greatly increase the strike rate.

\pagebreak

\section{Acknowledgement}
This work was made possible thanks to the support from the Department of Physics at Swarthmore College, as well as Prof. Lynne Molter from the department of Engineering for her valuable feedback throughout the research process. Author Simon Ji would also like to thank the Swarthmore College bowling club and his bowling teammates from Shanghai, China for unintentionally becoming the original inspiration of the project.

\section{Declarations}
The authors did not receive support from any organization for the submitted work.

\bibliographystyle{abbrv}

\end{document}